# Gas Emission Measurement System from Chilla-Juliaca -Perú

J. Mendoza Montoya, A. Olsson, S.- G. Mårtensson, K. J. Huanca Zea, A. R. Rojas Calla and J. Chilo

*Abstract*—In many countries around the world, most of the waste use to be disposed of to landfills, this generate public concern about the health effects of emissions. Landfill gases are produced by the natural bacterial decomposition of waste and it is about half of methane, with the remainder mostly carbon dioxide and minor amounts of other gases. Real-time measurement and modelling of emissions gases in landfills is important. In this work a low cost wireless measurement system is developed using MOS gas sensors (MQ4-1,MQ4-2, MQ5 and MQ9), microcontroller and XBee/HC-12 wireless communications modules. The system can be mounted on an unmanned aircraft (UAV, drone) or deployed as a wireless sensor network. Experiments have been carried out near a closed landfill, which show high gas concentration.

## I. Introduction

WE generate millions of tons of solid waste per day and it will increase because population and many economies are also growing continuously. Landfills are filled quickly. Landfills are a problem because they release methane ($CH_4$), carbon dioxide ($CO_2$) and other gases which are potent gases that can harm the health of the exposed population [1]. Understanding and measuring this kind of gases is very important to inform people living close landfills and the corresponding authorities.

Peru is an upper-middle-income country with economic growth in recent years. As a result, there is a growing middle class, increasing urbanization and increasing consumption that lead to increasing amounts of waste. The majority of solid waste is being dumped in unmanaged landfills without adequate treatment, leading to health and environmental problems [2].

There are major differences regarding waste generation, composition and management between developed and developing countries [3]. In Sweden, to name a developed country, the number of closed landfills, containing organic waste is estimated to be between 4000 and 8000. Most of them established during the last 100 years, thus still within their methane and $CO_2$-producing phase which is expected to be 100 years. To mitigate pollution originating from landfills Sweden prohibited deposition of organic waste in 2005 [4]. In developing countries as in the case of Peru it will take many years to take similar decisions.

Recently the authors in [5] have presented as estimation of methane generation rate in a landfill site which was computed as 0.012 per year and the methane generation capacity as 54.26 m3 /Mg. The authors have not found much written about measurement of gases emitted in landfills, however these measurements are performed in the same way when measuring greenhouse gases. Laser-based sensors for horizontal and vertical profiling of greenhouse gas levels in the atmosphere including an auto-aligning are presented in [6] a system that could be used for landfills gas monitoring. Polese et al. [7] have developed a wireless sensor networks for greenhouse monitoring, using commercial and ad hoc devices as layered double hydroxides sensors. With this system is possible to monitoring $CO_2$, NOX and other gases present in the greenhouse together with humidity, temperature and light intensity. A field campaign was carried out to measure the effect of reducing irrigation level on GHG emissions from soil and on crop yield on maize cultivated in Tuscany region (Italy) [8]. Monitoring was carried out using a transportable instrument developed by West Systems which is equipped with two detectors, one for $CO_2$ and $CH_4$ and the other one for $N_2O$.

As we see from above studies, greenhouse gas sensors and measurement methods should be considered. With the rapid development of electronics new and cheap gas sensors can be found in the market. WSN technology can be used as a measurement method, but it should be inexpensive and easy to install.

We have developed a low cost wireless measurement system to measure Carbon Dioxide ($CO_2$) and Methane ($CH_4$), using MOS sensors (MQ4, MQ5 and MQ9), microcontroller and XBee/HC-12 wireless communication modules and the interface to the computer is written in Python. The system could be deployed as wireless sensor network or mounted on an unmanned aerial vehicle (UAV, drone).

This paper is organized as follows. Section II describes the methods and materials used and describes the measurement

J. Mendoza Montoya, K. J. Huanca Zea and A. R. Rojas Calla are with Dept of Electrical Engineering, Universidad Andina Néstor Cáceres Velásquez Juliaca - Perú (jvaier.mmom@uancv.edu.pe, kevin.huancazea@uancv.edu.pe, ayrton.rojascalla@uancv.edu.pe).

A. Olsson is with the Dept. of Health and Caring Sciences, University of Gävle, Sweden (annakarin.olsson@hig.se).

S. – G. Mårtensson with the Dept. of Industrial Development, IT and Land Management, University of Gävle, Sweden (stig-goran.martensson@hig.se).

J. Chilo is with the Dept. of Electrical Engineering, Mathematics and Science, University of Gävle, Sweden (jco@hig.se).

setup. Section III introduces the measurement results and analysis. Finally, Section IV presents our conclusions.

## II. MATERIALS AND METHODS

Monitoring landfill gas emissions system is described in the following steps: hardware design, wireless transmission, software develop and measuring procedures.

### A. Hardware design

Figure 1 shows the circuit for four gas sensors (MQ4-1, MQ4-2, MQ5 and MQ9), where it can be seen that the output signal is connected to the ADS1115 module (16 bit ADC), the response is sent to Arduino Mega via I2C serial communication. LM358 operational amplifier does signal conditioning to make the signal from gas sensors suitable for data acquisition processing. The sensor requires two voltage inputs: heater voltage ($V_H$) and circuit voltage ($V_{cc}$). The heater voltage ($V_H$) is applied to the integrated heater to maintain the sensing element at a specific temperature, which is optimal for sensing. Circuit voltage ($V_{cc}$) is applied to allow measurement of voltage ($V_O$) across a load resistor ($R_L$), which is connected in series with the sensor.

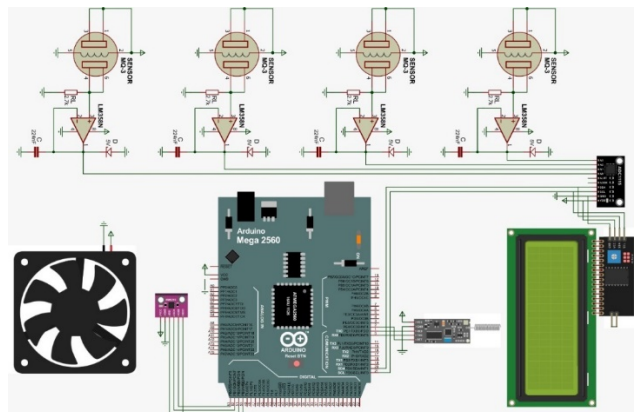

Fig. 1. Sensor circuit for MQ4-1, MQ4-2, MQ5 and MQ9, amplifier conditioning, connection to the Arduino analog inputs and the HC-12 module.

We have designed and built six modules that can be used to form a WSN. Figure 2 shows a finished module.

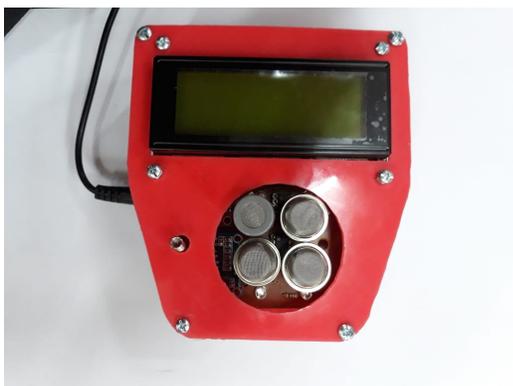

Fig. 2. Finished module with MQ4-1, MQ4-2, MQ5 and MQ9 gas sensors.

### B. Wireless transmission

The modules used in this work are HC-12. The HC-12 is a multi-channel embedded wireless data transfer module. Its wireless operating frequency band is 433.4-473.0 MHz This standard is configured for a 500m long distance transmission with a wireless receiving sensitivity of -112dBm as follow:

```
AT+FU3    (max long distance 600m)
AT+9600   (9600bps, 8 dibit data, no check, one stop bit)
AT+C001   (CH001 433.4MHz).
AT+P8     (20dBm, 100mW)
```

We have also evaluated three different DIGI XBee modules: XBee 802.15.4, XBee S2C ZigBee and XBee S3B DigiMesh, see Figure 3. These standards are designed specifically for energy efficient communications in a point-to-point configuration and includes sleeping and security.

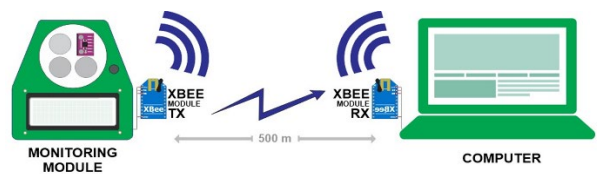

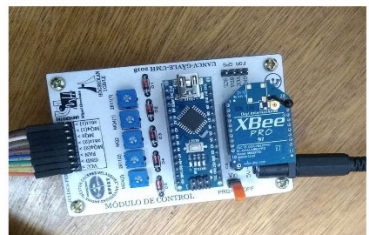

Fig. 3. Wireless transmition with Xbee module.

### C. Software

The developed software integrates two modules: the first module is written entirely in python. Python has a great advantage, it is based on open source code and has many libraries to improve data processing and graph plotting. The first module is designed to send and receive real-time data between sensor modules and the computer at 115200 bps. Data encryption is also coded to minimize loss data during transmission.

The second module is coded in Arduino-C language to be embedded in the microcontroller. The sensors detect traces of methane and carbon dioxide and the correlation voltage-ppm are shown according to the following equations:

$$ppm(ch4) = 152.4(\tfrac{2.2}{2.1}(\tfrac{4.8}{VADC} - 1))^{-2.71}, \qquad (1)$$

$$ppm(ch4) = 1000(\tfrac{2.2}{1.22}(\tfrac{4.73}{VADC} - 1))^{-2.807}, \qquad (2)$$

where $VADC$ is voltage from ADC device. Equation (1) is used to calculate methane ppm using data from MQ5 and equation (2) using data from MQ4.

III. MEASUREMENTS IN JULIACA-PERÚ

*A. Air pollution measurements in Juliaca city*

Juliaca city is one of the largest cities in Peru with an estimated population of more than 278,444 people, according to report given by the Instituto Nacional de Estadistica e Informatica. Nowadays, Juliaca City has serious environmental pollution problems, very high emissions of CO2 from vehicles. Another problem is that people cannot handle the garbage properly, which means that the CH4 emissions increase every year. Figure 4 shows a constant problem with the collection of garbage that causes waste dumps in all the city.

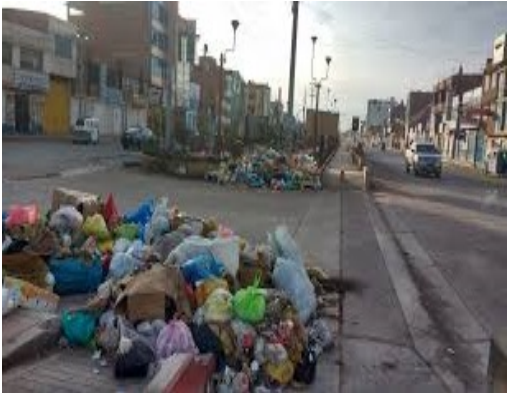

Fig. 4. Garbage in Juliaca city center.

Measurements have been carried out on six critical points in Juliaca city:

- Trade center (A)
- Huaynaroque mountain (B)
- Downtown street (C)
- Central Public Market (D)
- Main square (E)
- Andina University (F)

Figure 5 shows distance between trade center and all points of monitoring.

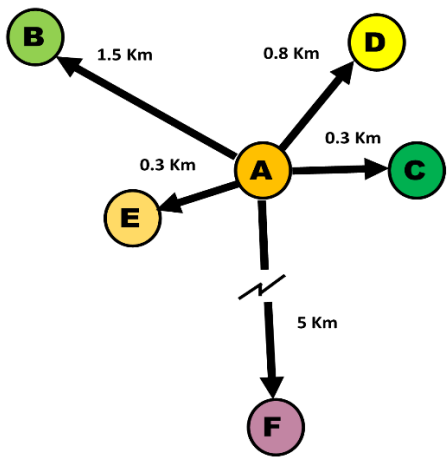

Fig. 5. Distances to trade center from monitoring points.

Comparative results of CH4 and CO2 for the six different places are shown in Figure 6. Methane concentration is going up to 3 ppm in places with high concentration of cars and population, but in places outside of the city, the concentration of methane is under 2 ppm. Concentration of carbon dioxide is higher in central public market, methane goes up to 700 ppm and the lower concentration is shown outside of the city with an average of 400 ppm.

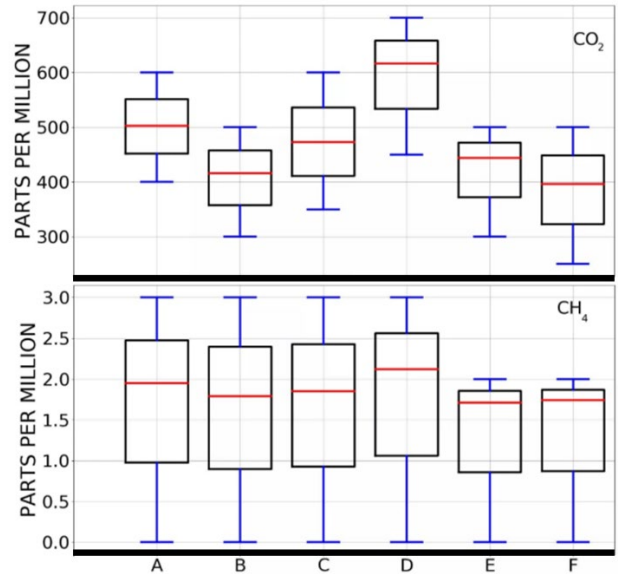

Fig. 6. Comparative results of CH4 and CO2 in Juliaca city.

*B. Air pollution measurements in Juliaca city*

In recent years, Juliaca population has increased considerably, which has generated uncontrolled increase of solid waste as consequence. More than 90% of the total waste generated daily by the population of Juliaca was deposited in the Chilla-Juliaca landfill [9]. In Figure 7, upper, we can see how the landfill has been covered with soil. Figure 7, lower, shows the landfill location, at a short distance you can see inhabited houses.

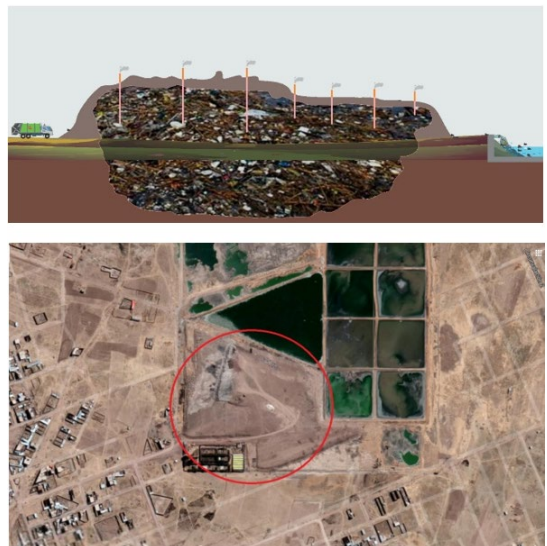

Fig. 7. Covered Chilla landfill (upper), Chilla landfill location.

According to the OEFA [10], in Peru there are more than ten landfills that do not meet the appropriate conditions for its operation, among them is the Chilla - Juliaca landfill. The situation worsens since there are more than 900 families settled around the dump, risking their health.

Many years there has been a problem with Chilla dump when it was open. One solution applied in those years was to burn ignitable garbage, the incineration of waste generates the production of all types of toxic pollutants that causes various diseases [11]. Given this situation, the authorities decided to close the landfill according to the following plan: 4 cells would be built, cover the solid wastes with soil, chimney installation and leachate drains building.

Methane measurements in two different locations are shown in Figure 8 and 9.

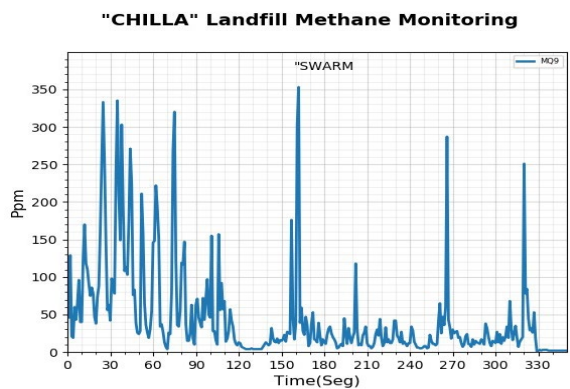

Fig. 8. CH4 monitoring in swarm place.

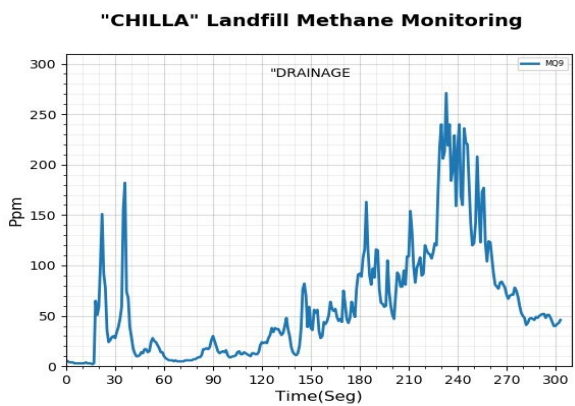

Fig. 9. CH4 monitoring in drainage place.

## IV. Conclusion

In this work we have designed cheap modules to measure polluting gases using MOS gas sensors microcontroller, XBee and HC-12. These modules have been tested in a city showing environmental pollution and in a landfill with high methane concentration.

According to the preliminary results, high levels of methane concentration can be seen. Keep in mind that the population is very close to the landfill, therefore it is very important to have more knowledge of the generation of methane and other gases. Future research will be aimed installing a WSN connected to the internet, IoT; additionally it will be used in drone to scan the movement of gases and plot it in 3D.